\newif\ifabstract
\abstracttrue
\abstractfalse 
\newif\iffull
\ifabstract \fullfalse \else \fulltrue \fi

\documentclass[11pt]{article}
\usepackage{amsfonts}
\usepackage{amssymb}
\usepackage{amstext}
\usepackage{amsmath}
\usepackage{amsthm}
\usepackage{algorithm}
\usepackage{algorithmic}
\usepackage{tikz}
\usepackage{multicol}
\usetikzlibrary{calc}
\usetikzlibrary{arrows,shapes}
\usepackage{dot2texi}
\usepackage{siunitx}
\usepackage{booktabs}
\usepackage{pdflscape}
\usepackage{threeparttable}
\usepackage{longtable}
\usepackage{multirow}
\usepackage{blindtext}
\usepackage[inline]{enumitem}
\usepackage{xcolor}

\usepackage{xspace}
\usepackage{theorem}
\usepackage{graphicx}
\usepackage{url}
\usepackage{graphics}
\usepackage{colordvi}
\usepackage{colordvi}
\usepackage{lscape}
\usepackage{natbib}
\usepackage[english]{babel}
\usepackage{float}
\usepackage{caption}
\usepackage{subcaption}
\usepackage{geometry}
\usepackage{amsmath}

\advance \oddsidemargin by -0.8in
\newcommand{\myparskip}{3pt}
\parskip \myparskip

{\hspace*{\fill}$\Box$\par\vspace{4mm}}

\begin{document}

\title{A Perspective on the Challenges and Opportunities for Privacy-Aware Big Transportation Data \footnote{\textbf{Accepted in the Journal of Big Data Analytics in Transportation}}}
\author{Godwin Badu-Marfo  \thanks{TRIP Lab, Concordia University, Email: {godwin.badu-marfo@mail.concordia.ca}} \and Bilal Farooq \thanks{Laboratory of Innovations in Transportation (LITrans), Ryerson University, Email: {bilal.farooq@ryerson.ca}} \and Zachary Patterson \thanks{TRIP Lab, Concordia University, Email: {zachary.patterson@mail.concordia.ca}}}

\begin{titlepage}
\maketitle

\thispagestyle{empty}

\begin{abstract}
In recent years, and especially since the development of the smartphone, enormous amounts of data relevant for transportation have become available. These data hold out the potential to redefine how transportation system (i.e. design, planning and operations) is done. While researchers in both academia and industry are making advances in using this data to transportation system ends (e.g. information inference from collected data), little attention has been paid to four larger scale challenges that will need to be overcome if the potential for Big Transportation Data is to be harnessed for transportation decision-making purposes. This paper aims to provide awareness of these large-scale challenges and provides insight into how we believe these challenges are likely to be met.
\end{abstract}

\end{titlepage}

\section{Introduction}
Transportation system (i.e. design, planning and operations) has been a quantitative discipline highly dependent upon data at least since the birth of modern travel demand modeling in the 1950s. Until recently, data collection has been done through dedicated, often self-reported surveys (e.g. household surveys, on-board surveys, etc.), and through various methodologies and technologies concentrated on vehicle flow counts (e.g. loop detectors). Recently, a combination of devices and technologies have dramatically increased the number of potential sources, as well as the amount of data that can be collected with urban transportation system applications, what we refer to as Big Transportation Data. Examples of this data include Bluetooth and CCTV traffic counts \cite{barcelo2010travel, cathey2005novel}, pedestrian counts with WiFi \cite{poucin2018activity, danalet2014bayesian, farooq2015ubiquitous}, activity detection with social media location data \cite{yazdizadeh2018automated}, dedicated travel survey smartphone applications \cite{patterson2017itinerumtrip} and smartphone data aggregators \cite{streetlight2018web}.

The potential for this data in transportation systems have not been overlooked, with many researchers in academia and the public and private \cite{lv2015traffic, dong2015traffic, zheng2016big, chen2016promises} sectors investigating ways in which to use it in their processes. Until now, the academic literature has been primarily preoccupied with two aspects of big data in transportation. First there has been research on how to go about collecting relevant data with these new technologies (e.g., \cite{leduc2008road},\cite{shi2015big}, \cite{patterson2017itinerumtrip}). Second, there has been research focusing on methods (statistical, machine learning, etc.) using collected data and inferring transportation relevant information from it (e.g. mode, trip purpose, etc.) \cite{yazdizadeh2018automated, nitsche2014supporting, zhang2014high}.

While the successful collection of data, and inference of information relevant to transportation system presents many challenges to the routine incorporation of Big Transportation Data in design, planning and operations, little attention has been paid to the impending challenge of actually being able to store, manage and process all the data on large and operational scale, not to mention the challenge of protecting privacy of the people providing the data. We divide these large-scale implementation challenges into four dimensions. The tautological fact that there is a large quantity of Big Data presents challenges in storing it. Second, the need to compute algorithms on large scale data presents a challenge in processing. Third, Big Data comes in many different formats, making the ability to take advantage of data collected from different sources challenging. Fourth is the challenge of protecting personal privacy.

While the quantity of data and diversity of formats are primarily technical challenges, personal privacy is a political as well as technical challenge. The political nature of the challenge was recently evidenced by the controversy around Facebook and Cambridge Analytica \cite{olivia2018facebook} and public reaction to it. The issue of privacy and Big Data is multifaceted. Most obviously, much Big Data is sufficiently detailed (e.g. geographically and temporally precise GPS data) that it could reasonably be used to identify individuals. A less obvious challenge to privacy is the ability to combine information about individuals across data sources thereby making the identification of individuals possible with individual ``quasi-identifying" information. Another less obviously personal challenge relates to who can access private data, and how to control access in the most secure way. 

All of these challenges will need to be met before the potential for Big Data in transportation can be harnessed. As such, this paper aims to provide an in-depth awareness of the large-scale implementation challenges currently facing the use of Big Transportation Data in design, planning and operations of transportation. It also  provides insights into how we believe these challenges are likely to be met.

The paper continues with a section describing the scope of this paper and moves on to define Big Data, Big Transportation Data and from where they come. The next section describes the current state of the transportation literature as it relates to Big Data. This is followed by a background section on system architecture needed to understand the sections on the four main challenges to the widespread use of Big Transportation Data in transportation planning. A concluding section sketches our understanding of how the challenges of Big Transportation Data are likely to be overcome in the future.

\section{Scope of this Work}
The four large-scale challenges to the widespread use of Big Transportation Data identified in this paper have resulted from a thorough literature review. Since there is very little attention to this question in the transportation literature, most of the literature reviewed has come from computer science, computer engineering, and fields the most advanced in the use of Big Data, such as health and agriculture. The primary Google Scholar search terms used were``big data implementation challenges" and ``big data technologies." Relevant papers from articles resulting from these searches were then included in the literature, and this process was done iteratively. The more than one hundred and fifty papers resulting from this process were placed into four categories of challenges: storage, processing, integration, and data privacy. These challenges concentrate on those relating directly and uniquely to Big Data. While other challenges such as data security, integrity and transfer are relevant to Big Data, they are not unique to Big Data, and so we don't concentrate on them here. Interested readers can consult the vast literature on these topics elsewhere \citep{tankard2012big, tierney2012efficient, lagoze2014big}. We continue by defining both Big Data and Big Transportation Data, as well as from where they come.

\section{Key Characteristics of Big (Transportation) Data}
Big Data has been described, characterized and defined in both academic and non-academic (traditional media, trade press, etc.) sources. Across these sources, there is a great variety in how Big Data has been defined and characterized \cite{mcafee2012big, hashem2015rise, zikopoulos2011understanding, wu2014data}. Often, Big Data are characterized by words beginning with the letter ``v." One problem with such``v-words" is that there is often variation in how they are defined from one author to another. Also,``v-words" do not necessarily define characteristics of only Big Data but of ``non-Big-Data" as well. Finally, there are some concepts critical to understanding the challenges for the widespread use of Big Data that are not easily described with``v-words." Given the confusion around definitions and the fact that we are most interested in the characteristics of Big Data as they relate to the challenges of using it, we discuss two types of characteristics, not all of which are ``v-words." As such, below we discuss ``defining" and ``non-defining" characteristics of Big Data.

\subsection{Defining Characteristics of Big Data}
Defining characteristics of Big Data are those that are unique to Big Data as opposed to data in general. Those critical to understanding the challenges of widespread use of Big Data are those from the most-cited definition of Big Data by the Information Technology (IT) advisory firm Gartner. According to Gartner:

\begin{quote}
	``Big data is high-\textit{volume}, high-\textit{velocity} and/or high-\textit{variety} information..."\cite{gartnerwebsite}.
\end{quote}

\textbf{Volume} refers to the size of individual datasets. Already in 2011, there were 2.5 quintillion bytes of data created every day \cite{hilbert2011world}, and this number keeps increasing exponentially \cite{jagadish2014big, kahn2011future} so that``Big" datasets currently typically range from zettabytes ($10^{21}$ bytes) to yottabytes ($10^{24}$ bytes) \cite{chen2014data}. It is often said that ``Big" datasets are too large to be handled by an individual computer \cite{katal2013big}. 

\textbf{Velocity} refers to the rate at which data are being generated. As with volume, the figures on rates of data being produced and received can be staggering. It was reported in March 2018 that over 900 million photos were  uploaded to Facebook \cite{gewirtz2018volume}. In addition to being related to the rate with which data are generated, velocity also encompasses a notion sometimes referred to in the literature as \textit{variability} \citep{gandomi2015beyond}. Whereas velocity refers to the rate at which data are generated, variability refers to variance over time in data flow rates. 

\textbf{Variety} refers to the structural heterogeneity of data. That is, data provided in different formats, some structured and others not. Structured data are mostly in the form of tabular schema-imitating spreadsheet and relational database systems. Text, audio, images and videos are examples of unstructured data, with 
Extensible Markup Language (XML), being an example of semi-structured format \cite{gandomi2015beyond}. Unstructured data is more difficult to process, store and integrate and is becoming more common \cite{mansuri2006integrating, choi2006real, doan2009information}.

\subsection{Non-defining Characteristic of Big Data}
A non-defining characteristic of Big Data is simply one that applies to other types of data as well. Such characteristic creates an important challenge for the widespread use of Big Data, as Big (and non-big) Data is either by nature personal, or can be personal. By personal we mean that an individual's identity is explicit, or can be revealed. That data \textit{can} be personal we mean that different data sources can be combined to identify an individual and other information about them. While this is not a new problem (e.g., it has been a concern for a long time with traditional census data \citep{samarati2001protecting}), it becomes compounded with Big Data. This is so because of the many potential different sources of data available on people \cite{doan2009information} and also because of the very personal nature of some Big Data (e.g. precise location data, medical records, etc.) \cite{xu2014information}.

Recently, large data collection organizations (i.e. government, institutions and non-governmental organizations) have begun adopting ``open data"" initiatives that allow for data to be freely available, shared, redistributed and reused by the public without restrictions of use \citep{auer2007dbpedia}. As such, open data can serve as a resource for private, public and academic research. The availability of such data means privacy has become of even greater concern.  

\subsubsection{Big Transportation Data}
We characterize Big Transportation Data (BTD) simply as Big Data (as characterized above), but with potential transportation system applications. That is, data that could be used in areas in the traditional purview of transportation design, planning and operations, such as travel demand forecasting, infrastructure planning, transit network planning, operation optimization,etc. 

\section{Where Does Big Transportation Data Come from?} \label{sec:ecosystem}
BTD comes from the combination of three types of technologies. We begin with two broad categories of devices that collect BTD; location-ignorant and location-aware devices. Location-ignorant devices are able to sense the presence of other devices, although they are not explicitly aware of their own locations. These include technologies such as Bluetooth\citep{perego2017wireless}, Wireless Fidelity (WiFi)\citep{perego2017wireless}, Global System for Mobile (GSM)\citep{perego2017wireless} and Closed-circuit Television (CCTV)\citep{gadhe2017networking}.  

\begin{figure}[h]
	\centering
	\includegraphics[width=1.0\textwidth]{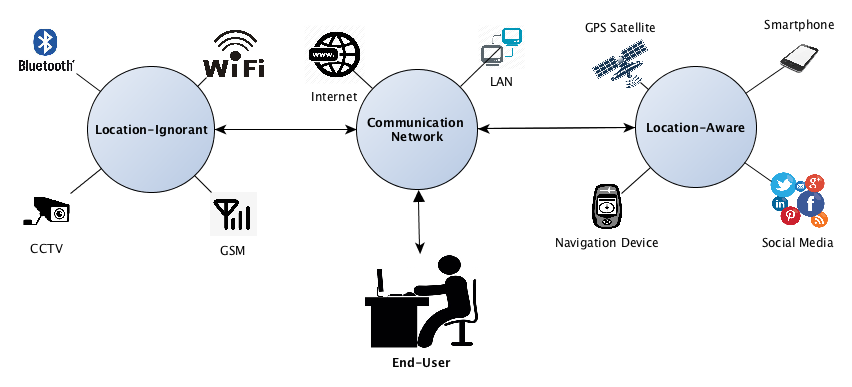}
	\caption{Ecosystem of Big Transportation Data}
	\label{fig:btdEcosystem}
\end{figure}

The second are devices that can determine their own whereabouts, i.e. they are location-aware. These devices typically derive their locations based on the location of other devices such as WiFi routers, GSM towers, or satellites part of various Navigation Satellite Systems, such as the Global Positioning System (GPS). They include GPS units, GPS navigators and most importantly smartphones. 

While devices that collect data are critical for being able to use BTD, its potential can only be harnessed if the devices are connected to a communications network, such as the Internet\cite{stamp2011information}, private Local-Area Networks (LAN)\cite{stamp2011information} or Wide-Area Networks (WAN)\cite{stamp2011information}. These networks allow the transfer of data from collecting devices to database storage systems from where they will be accessed for processing and analysis by end-users. Figure \ref{fig:btdEcosystem} provides a schema of the BTD Ecosystem.

\section{The Current State of BTD in Transportation}
The combination of location-ignorant, location-aware and communications networks has led to the birth of Big Transportation Data. Academia as well as the public and private sectors have not overlooked the potential for BTD in transportation. 

\subsection{Research with Data Collected with Location-ignorant Devices}
In recent years, academic research has been conducted with the use of data from location-ignorant devices in public transit planning and operations. Transit smartcard data has been at the forefront of this to understand travel behavior \cite{bagchi2005potential,pelletier2011smart} and transit user loyalty \cite{trepanier2010assessing}, also state that Smart cards can be used to ascertain the loyalty of transit users in a network. 

WiFi network data has also been used to understand (primarily pedestrian) travel behavior based on connection histories to wireless routers \cite{poucin2016pedestrian,shlayan2016exploring}. Similarly, Bluetooth receivers have been used to assess automobile route choice and travel times on alternate routes \citep{hainen2011estimating}.

\subsection{Research with Data Collected with Location-aware Devices}
Location-aware technologies have been developed to determine their own location. Location sensors derive precise locations through the use of GSM, WiFi and GPS \citep{leick2015gps,van2009gps}. Transport operations, planning and research heavily rely on these devices for precise spatio-temporal data in analysis and decision making. Location-aware technologies is discussed in two categories namely GPS and Smartphones.

\subsubsection{GPS}
Navigation GPS devices have long been used for finding the location of Point Of Interest (POI). Transportation fleet operations rely heavily on navigation GPS systems that provide mobility trajectories of fleets. Much academic research has been done to cover the application of navigation GPS devices in transportation. Davies et al.\citep{davies2010evaluating} evaluated the use of GPS devices for providing location-aware visual and auditory prompts for people with intellectual disabilities to enable them in navigating busroutes. Handheld GPS devices have been extensively used for travel mobility surveys in research \citep{draijer2000global, stopher2007household, montini2015comparison}. A study on children's mobility using GPS-tracking device and mobile phone survey was conducted in Copenhagen \citep{mikkelsen2009children}. The research shown diversity of mobility patterns for children and the geographic interdependency of child mobility. Surveying and data collection with Navigation GPS devices are becoming phased out due to the emergence of location-aware Smartphones, that assure precise location from satellites and can augment location from cell phone towers in places with poor satellite signals.

\subsubsection{Smartphones}
Pervasive Smartphone devices have gained popularity recently for mobile and internet communication. Many mobile applications (e.g. social media, maps, dating apps, locations and others) are used daily on smartphones by their users. Location-aware applications are common in smartphones, they observe the location of the user and report to a location based service (LBS). Location Based Services provide queries of point of interest within a defined proximity of the user as reported by the smartphone. As an example, a Smartphone user can ask (query) for restaurants near-by or within a distance of his/her current location to receive a list of matched restaurants. Smartphones have inbuilt Assisted-GPS sensor for precise location tracking to satellites, in cases where cloud visibility is achieved. At places with less cloud visibility, Smartphones can gain location by connecting to nearest cellphone towers or WiFi access points. A large body of literature has contributed to the use of Smartphone in transportation studies. 

Patterson et al. \citep{patterson2016datamobile} conducted an experiment on participants from Concordia University, that used a smartphone travel survey developed to collect passive data on human mobility whilst minimizing the respondent burden. Respondent burden is reduced in such surveys relative to traditional self-reported surveys. An enormous amount of location-sensitive data is gathered on social media platforms like Facebook, Twitter, Instagram and others.

\subsubsection{Information Inference from BTD}
Another research area receiving attention in the transportation literature is that related to the development of methods allowing the inference of the main aspects of transportation demand required for traditional trip-based transportation demand forecasting. As such, data inference methods have been developed in the following areas. The inference of trip ends was one of the earliest questions to be broached in the literature (e.g. \cite{stopher2007household}), but one which continues to have (primarily rule-based methods) methods developed (e.g. \cite{zhao2015stop,patterson2016datamobile}). 
Mode detection has received the greatest amount of attention in the literature with methods evolving from rule-based (e.g. \cite{bohte2009deriving}) to discrete choice (e.g. \cite{bierlaire2013probabilistic}) and machine-learning approaches \cite{gonzalez2010automating,reddy2010using,gonzalez2010automating}. 

Purpose detection, has turned out to be the most difficult to infer. Initial rule-based (e.g. \cite{wolf2001elimination} continue to be used (e.g. \cite{shen2013process}) but are being replaced with machine learning algorithms(e.g. \cite{mcgowen2007evaluating,griffin2005decision}) increasingly using data collected from various BTD sources such as social media (e.g., \cite{yazdizadeh2018automated}).

Finally, itinerary inference has evolved from simple map matching methods (see \cite{white2000some}) to more sophisticated probabilistic approaches \cite{bierlaire2013probabilistic}. Itinerary inference have been applied primarily to road networks and particularly to automobiles (e.g. \cite{bierlaire2013probabilistic}) and bicycles (e.g. \cite{hood2011gps}). Less common are methods for inferring transit itineraries combining smartphone and GTFS data \cite{zahabi2017transit}.

\subsubsection{Future sources of BTD}
In addition to current sources of BTD, we also have to include the coming addition of autonomous vehicles as a data source. According to Intel\cite{intel2016data}, the evolution of Autonomous Vehicles (AV) with their on-vehicle sensors and cameras will generate and require enormous amounts of data. AV cameras alone will generate 20 to 40 Mbps per vehicle, while radars will generate between 10 and 100Kbps with an estimated average of 40 terabytes of data for every eight (8) hours of driving \cite{intel2016data}.

\subsubsection{Summary of Current BTD Research}
As can be seen from the rest of this section, there is a great deal of research being done on BTD in transportation. Collectively this work can be divided into three broad categories. The first category relates to the use of various technologies in actual data collection \cite{arentze2000data, efthymiou2012use}. The second category concentrates on challenges related to methods that process BTD and seek to infer information from it that can be useful in transportation \cite{yazdizadeh2018automated}. The third category focuses on the evolving technologies that present opportunities for the successful implementation of BTD. While this work is clearly necessary for BTD to be effectively used in transportation, there has been little emphasis on the importance of system architectural components necessary for large-scale adoption of BTD. 

\section{System Architectural Components}
Critical to understanding the challenges of BTD is an understanding of data system architecture more generally. Data Management Architectures (DMAs) organize the flow of data from collecting devices to the storage systems with which data is managed. DMAs can be split into three essential elements. First is the physical infrastructure (i.e., hardware) needed to be able to store data. Second are file systems with which files (and their underlying data) are organized on hard drives. Third are database management systems.

\begin{figure}[h]
	\centering
	\includegraphics[width=1.0\textwidth]{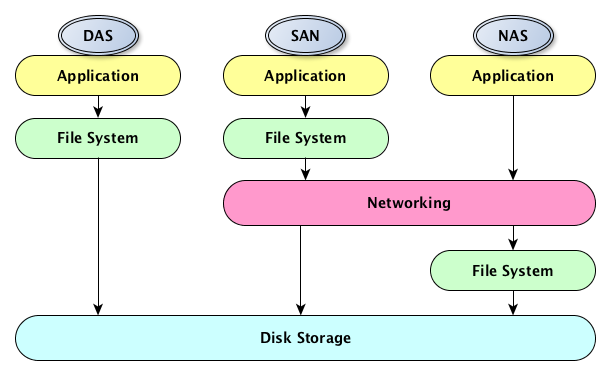}
	\caption{Disk Storage Drives}
	\label{fig:storagedevice}
\end{figure}

\subsection{Hardware}
We begin with the hardware side of data management systems and with data retrieval. Data retrieval typically, and traditionally, involves an in-between step; data must be read from long-term storage on hard drives into active memory. The speed with which this happens is dependent upon three elements: computer processor (CPU), disk characteristics, and disk connection to active memory. The faster the processor, the faster data can be read into active memory \citep{adabi2018optimizing, ousterhout1989beating}. Disks themselves vary in the speed with which data can be accessed from them. Traditional spinning Hard Disk Drives (HDDs) have slower transfer speeds than Solid State Drives (SDDs), from which data can be accessed directly from its storage sector \cite{tsirogiannis2009query}. Finally, the connection between hard drives and active memory plays a critical role in the speed with which data can be accessed, See Figure \ref{fig:storagedevice}. Transfer speeds are fastest from directly attached storage (DAS) (i.e. hard-drive on a single node, such as a server or other standalone computer). Speeds decrease with a greater separation of where the data is stored and active memory with network attached storage (NAS) (i.e. connected through a local area network (LAN) having slower speeds than DAS, and storage area networks (SANs) (e.g. storage on remote networks) potentially taking even longer than LANs \citep{adabi2018optimizing, patil2016study}. The writing of data to storage involves the reverse process, i.e. from active memory to final storage.

\subsection{File Systems} \label{sec:fileSystems}
On hard drives, data is stored hierarchically. At the lowest level, data is stored in a binary format as bytes with a location on a hard drive \citep{ousterhout1989beating, adabi2018optimizing}. Bytes are grouped together as ``data'' (e.g. the content of a spreadsheet cell) and data are grouped together into files. There are different underlying logical systems by which bytes can be organized into data, and data into files. These logical systems are known as ``file systems," that are a subsystem within the operating system (e.g. Linux, Windows, MacOS, etc.) \citep{tanenbaum1987operating, ousterhout1989beating}. There are many file systems that exist, but the most common are NTFS, VFAT, EXT3 and HPFS \citep{tanenbaum1987operating}. 

\subsection{Database Management Systems} \label{sec:RDBMS}
While file systems hierarchically organize data and files on hard drives, database management software uses the file system to make data available for processing. This is done with database software. The traditional and most popular database software products are based on Structured Query Language (SQL). SQL resulted from the work of E.F. Codd who introduced the``Relational Model" in the 1970s \citep{codd1970relational}. As a result, these products are also known as Relational Database Management Systems (RDBMS) of which there are many examples (e.g., MySQL, PostgreSQL, Microsoft SQL Server, Oracle DB). RDBMSs, now typically referred to as``legacy" systems, have proven very efficient for intensive amounts of data storage, retrieval and processing for many decades \citep{vicknair2010comparison}. RDBMSs are organized into databases containing tables, with tables related to each other by common identifier constraints (i.e., keys). Database table schemas are strictly defined. That is, data can only be read into them if it adheres to the structure defined in the schema (e.g. text data cannot be read into a variable defined as an integer). The structure placed upon the data is a primary factor making such systems so efficient at saving and accessing data. Also, RDMBSs are typically``centralized" meaning they are deployed on one node and cannot be easily scaled to multiple nodes.

Finally, RDBMSs are ``transactional" \cite{gray1992transaction, maier1983theory} which means they also demonstrate the following properties. First, Atomicity guarantees that all transaction operations are executed ``all-or-nothing"; if one part of a query fails, the entire query fails and none of it is executed. Second, transactional Consistency guarantees every transaction will bring the database from one valid state to another. Third, Isolation ensures concurrent transactions (e.g. from multiple users) will be executed sequentially. Fourth, Durability ensures that once a transaction has been committed, databases remain the same in the event of a power loss, system error, crash, etc. Collectively these four characteristics are known as ``ACID" properties of a transaction \citep{maier1983theory}.

\section{Challenges and Opportunities in ``Storing-It-All''} \label{storingItAll}
The first challenge identified in the literature is to actually being able to store and manage all the BTD. This concerns the ``v-word'' ``volume.'' The volume of data that will need to be stored is a challenge for using Big Data in general, but is clearly also a challenge in transportation in particular with the many new sources of data (described in Section \ref{sec:ecosystem}) available with transportation applications. As an example, it is now possible to record mobility traces collected by cell phone operators, traffic information, transaction systems (integrated ticketing, road user charging, car park payment, electronic fee collection), cameras, in-vehicle GPS, social media and smart phone geolocation technologies~\cite{chen2014data}. The rich data gathered from these sources will help to improve on transport modelling and planning to deliver accessibility, efficiency and economic performance potential which hitherto was not possible. Ultimately, this boils down to adding capacity; faster CPU, hard drives with more storage capacity from which data can be accessed (and written) more quickly, and enabling software. Such capacity can be added in two ways; vertically, or horizontally (see Figure \ref{fig:scaledsystem}). 

\subsection{Vertically Scaled Systems} \label{sec:vertical}
\begin{figure}[h]
	\centering
	\includegraphics[width=1.0\textwidth]{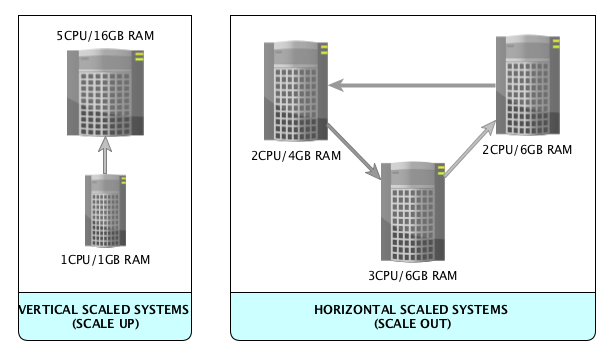}
	\caption{Scaled Systems (\textit{Systems sizes are for illustration purposes only})}
	\label{fig:scaledsystem}
\end{figure}

The traditional approach to increasing data storage and management capacity is ``vertical scaling". This involves improving the capacities of a single node (i.e., a standalone computer). Since traditional RDBMSs were designed for deployment on such systems, there are few software implications and as a result, vertically-scaling concerns primarily hardware. As such, it entails the use of faster CPUs, the increase of active memory (RAM) and the addition of larger and faster disk drives (e.g. converting from HDD to SSD) as shown in Figure \ref{fig:scaledsystem}. 

While hardware improvements lead to vertical scaling, there are limitations to just how ``high" such systems can be scaled. While Moore's law suggests increasing improvements in CPU speeds, we are limited to the available chip technology at any given time \citep{schaller1997moore, ousterhout1989beating}, even when considering the possibility of multiple cores on the same node. Secondly, there is no guarantee that Moore's law will continue into the future \citep{kish2002end, ousterhout1989beating}. Similarly, capacities are limited by available active and long-term storage technologies. Moreover, it may be possible to scale up to required capacity with available technology in some circumstances, but component cost increases dramatically with improvements at the cutting-edge of performance. Finally, vertically scaling a single node amounts to putting all of your eggs in one basket, the downside of which is that if there is a problem with the vertically-scaled node (e.g., it crashes), data cannot be read or written. In other words vertical integration increases the risk of greater downtime. 

\subsection{Horizontally Scaled Systems} \label{sec:horizontal}
Instead of increasing the capacity of a given node, horizontal``scaling-out" involves the combination of different nodes into a  ``cluster." That is, a ``distributed" storage system. As illustrated in Figure \ref{fig:scaledsystem},  nodes with similar (homogeneous) or varying (heterogeneous) capacities are added to the cluster to meet storage and computing needs. Distributed systems have the following advantages compared to single-node systems. First, it is possible to add resources (CPU, active and long-term memory) in a cost-effective manner since capacity can be increased almost limitlessly without the skyrocketing costs associated with performance increases in a single-node.

Second, distributed systems typically store redundant copies of data across multiple nodes, which decreases the risk of data not being available at any given time. The storage of multiple copies is done in the following ways. The same data can be stored on different nodes. This, referred to as ``redundancy," means that if one node goes offline, the data is still available on another node. Additionally, data can be be ``sharded." This means that different parts of the same dataset can be stored on different nodes. For example the columns (or rows) from the same database table can be stored separately, thus increasing the speed at which data can be accessed and written.

As with vertically scaled systems, database software is required for the proper functioning of horizontally scaled systems. At the same time, the limitations of traditional RDBMSs make them inappropriate for horizontally scaled systems. A key characteristic of horizontally scaled systems, is that data is synchronized across nodes within the system. Traditional RDBMSs were not initially designed with this in mind, so they remain relatively inflexible in this respect making synchronization with them inefficient and arduous \cite{moniruzzaman2013nosql, vaquero2011dynamically}. This inflexibility is ultimately due to the reliance of RDBMSs on traditional, centralized file systems (see Section \ref{sec:fileSystems}). Such file systems do not easily allow the management of files across multiple nodes.

As a result, horizontal scaling requires both hardware in the form of nodes and networks, as well as Distributed Database Management Systems (DDMS) that are designed to seamlessly synchronize data across nodes. In order to do this, DDMSs themselves rely on non-centralized distributed file systems. DDMSs and files systems make up the software component of horizontally-scaled systems \citep{vaquero2011dynamically}.

\subsubsection{Distributed File Systems} \label{sec:dfs}
The logical hierarchy of centralized file systems locates bytes on a single hard drive and groups the bytes into data and files. Distributed file systems on the other hand use a slightly deeper hierarchy. Bytes are stored on a hard drive, organized into data, data are organized into ``chunks" and chunks into files \citep{ousterhout1989beating}. Chunks themselves, however, do not have to be stored on the same hard drive. So, in addition to a deeper logical hierarchy, the key feature of distributed file systems is that they can also locate data across different hard drives. While several distributed file systems exist, the most common are the the Google File System (GFS) and the Hadoop File System (HDFS).

GFS, developed by by Google Inc. \citep{google2018online} supports large-scale and data-intensive applications \citep{ghemawat2003google}. It can be deployed on any standard node thus making it desirable from a cost perspective when scaling-out a system. The distribution of chunks across hard drives with GFS is orchestrated by one ``master" node to the subnodes (``slaves") of the system. This organization means that if the master node goes offline, access to data on the master and slave drives becomes impossible. As such, GFS is said to have a single point of failure.

The Hadoop File System (HDFS) \citep{shvachko2010hadoop}, designed by Apache like GFS also runs on any standard node and is suitable for data-intensive applications. It is also based on a ``master-slave" architecture, and as a result also has a single point of failure. Compared to GFS, HDFS has become much more common in industry application, and has had a series of DDMSs built using the underlying HDFS \citep{shvachko2010hadoop, borthakur2007hadoop, shafer2010hadoop}.

\subsubsection{Distributed Database Management Systems} \label{sec:ddms}
In addition to specialized file systems, and due to the limitations RDBMS, horizontally-scaling also requires dedicated database management systems (DDMSs). A number of such systems exist and fall into broad categories; structured and unstructured. Basically, such systems are distributed versions of RDBMSs. That is, they allow for the distribution and synchronization of data across multiple nodes, but they remain structured database management systems. The most common such systems in use are Google Big Table \citep{chang2008bigtable} and Apache HBASE \citep{vora2011hadoop}. Another increasingly common DDMS is noSQL, which in addition to being designed for horizontally-scaling is also unstructured. 

\subsection{Characteristics of Horizontally-Scaled Systems} \label{sec:charHorizontal}
In order to be effective, horizontally-scaled systems need to be planned well. Key characteristics of effective distributed systems have been summarized in Brewer's CAP Theory \cite{brewer2000towards, gilbert2002brewer} (see Figure \ref{fig:capTheorem}):

\begin{figure}[h]
	\centering
	\includegraphics[width=1.0\textwidth]{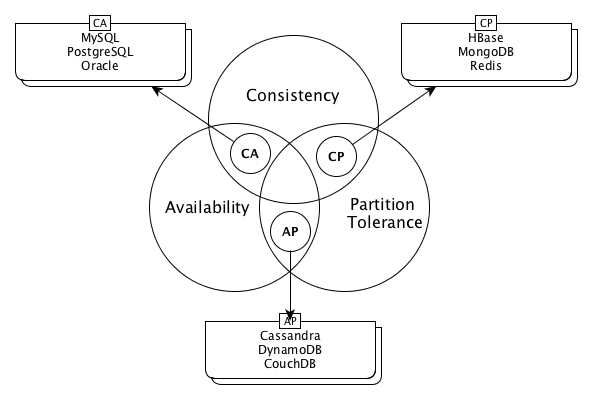}
	\caption{CAP Theorem}
	\label{fig:capTheorem}
\end{figure}

Consistency (C): While redundancy means having multiple copies of the same data in different locations,``Consistency" means that all copies of redundant data are identical \cite{oracle2015managing}. This ensures that the most up-to-date data is available even if there are server or network failures. 

Availability (A): Distributed Systems operate on multiple nodes that run concurrently in the implementation of a task. As a result, individual nodes can stop operating (e.g., due to a crash failure). Such failures are common and inevitable in networked systems. Availability means there is a sufficient number of nodes with redundant data that all data can be accessed at all times, even if one or multiple nodes crash \cite{oracle2015managing}.

Partition Tolerance (P): Partition tolerance is similar to availability in that it describes systems where redundant data can be accessed at all times. With Partition Tolerance, however, the concern is not with nodes themselves, but with the network connectivity of the nodes \cite{fathi2013integration}. This can be seen as``network availability."

While ideally, distributed systems would have all three of these characteristics, in practice they are typically characterized by two at most, with system design amounting to trading-off between the characteristics \cite{gilbert2012perspectives}. While systems that are not distributed over different networks exist, discussion on distributed systems is typically limited to those that are. As a result, we describe only systems demonstrating Partition Tolerance that is AP and CP systems.

AP systems are characterized by Availability and Partition Tolerance. Such systems are made up of multiple networks (P) with a node (or cluster of nodes) (A) on each network. Additionally each node (or cluster) would be able to operate without communication to the others. If communications between the nodes/clusters were interrupted, updates to data would be out of sync and as such, the system is not always consistent (i.e., it does not demonstrate Strong Consistency). Once all networks are functioning, data will become synchronized again but with delays (Eventual Consistency). Well known ``AP" systems include CouchDB and Cassandra.

CP Systems are characterized by Consistency (C) and Partition Tolerance (P). Such systems are made up of multiple networks (P), but with only single nodes on each network. CP Systems maintain multiple copies of the same data and therefore are ``Strongly Consistent." Unlike AP systems, if there is a network failure, there is always sufficient network redundancy, that the data across all nodes remains consistent. At the same time, since there is only one node per network if one of the nodes fails, there is no node redundancy, and as a result, the system is not ``Available." Well known ``CP" systems include MongoDB and Redis.

As such, the volume of BTD presents a major challenge to the potential to use it effectively in transportation in the future. At the same time, new approaches and technologies, namely the use of scalable distributed systems appear to be the most probable solutions to meeting this challenge, with the design of the systems requiring choices and trade-offs to be made between Consistency, Availability and Partition Tolerance. 

\subsection{Data Storage Opportunities for Transport Systems}
The recent advent of sensor-based technologies such as infrared detectors, video detectors, induction coils at bayonet points, laser detectors and others, for real-time traffic monitoring and the passive data collection of mobile user trip data for transport modelling (i.e. mode and activity inference) contribute a rich dataset for real time analytics and decision making by transport stakeholders. In this regard, Damaiyanti et al.~\citep{damaiyanti2014querying} presented a novel system that collects traffic data and represents speed values of all road segments of Busan. Their system stores traffic data and supports traffic congestion queries in a distributed NoSQL document database system that is deployed on a MapReduce framework. The rapid rate at which transport data is ingested in an ITS ecosystem, as earlier discussed, makes an adaptation of a distributed database system a requirement to achieve an effective and performing transport system. The United States Department of Transport \cite{usdot2013transport} has stated the data streams rate within 10 and 27 petabytes per second of connected vehicle Basic Safety Messages (BSM) will be generated, and thus that connected vehicle-to-vehicle (V2V) infrastructure is being implemented in test tracks. These implementations require a large volume of distributed data warehouse capacity. Amini et al.~\cite{amini2017big} proposed a comprehensive and flexible architecture based on a distributed computing platform for real-time traffic control. Using a mapReduce framework, their distributed architecture is based on systematic analysis of the requirements of an existing traffic control systems and analytics engine that informs the control logic.

\section{Challenges and Opportunities in Unstructured Data Storage} \label{sec:unstructuredDataStorage}
The second challenge identified in the literature is being able manage BTD of many different data formats. This concerns the v-word, ``variety." As with data volume, this is a challenge to using Big Data more generally, as well as BTD.

In general, data can be formatted on a continuum between structured and completely unstructured data. Structured data (described in Section \ref{sec:RDBMS}) is highly organized and format schema are defined before data is even collected (i.e., before it is stored in a database). In fact, if structured data is expected for a relational database but the data is not sent in the pre-determined format, it will typically not be stored at all. On the other end of the spectrum is unstructured data. Unstructured data is negatively defined as that not adhering to any predefined data schema. It comes in two main types; text and non-text. Examples of unstructured text data are email messages, text documents, etc. Examples of non-text unstructured data are satellite images, CCTV videos, etc. In between structured and unstructured data there also exists semi-structured data. Semi-structured data encapsulates unstructured data within a meta-structure using semantic tags and marking. Common semi-structured formats include mark-up languages  (e.g. HTML, XML) and JSON (Java Script Object Notations). 
Different formats present two major challenges. First, mechanisms are required to be able to save and access the data in an efficient manner. Recall that structure in traditional RDBDMSs is what allows them to efficiently manage large amounts of data. Second, taking advantage of BTD also means taking advantage of different sources of data, typically in different formats, so integration of the different data sources is a challenge. Being able to use data of different formats ultimately requires the use of software that can accommodate a variety of formats in a structured manner that also allows efficient retrieval. The most common DDMSs rely on frameworks based on NoSQL \citep{pokorny2013nosql, moniruzzaman2013nosql} with NewSQL being a more recent and quickly evolving framework.

\subsection{NoSQL and NewSQL} \label{sec:noSQLnewSQL}
NoSQL databases (i.e., non formally structured relational databases) are becoming more popular for big data storage. NoSQL databases are much more flexible allowing the following features that are impossible in RDBMSs: the ability to add new variables and modify existing variables within tables, without the need to drop and recreate tables; support for copying and pasting data into and from tables; more flexible integration of different programming platforms through Application Programming Interfaces (APIs); eventual consistency (see Section \ref{sec:charHorizontal}), and supports the management of data across nodes and in quantities too large for one node. At the same time NoSQL systems are not transactional, and as a result do not demonstrate ACID properties (see Section \ref{sec:RDBMS}). NoSQL databases are becoming the core technology for big data and can be characterized according to one of four data models:
key-value, column-oriented, document-oriented, and graph. We describe these models below.

In Key-value databases each observation (row) is stored as a dictionary, with each key defining a variable. Queries can be made directly according to keys. Such databases are characterized by high expandability (easy to add or remove variables without having to create new tables) and shorter query response time than those of relational databases. These databases have suitable storage structure for continuously growing, inconsistent values of big data for which faster response of queries is required. Key-value databases provide support to large-volume data storage and concurrent query operations. Popular examples of Key-value NoSQL DDMSs are MongoDB \citep{dirolf2010mongodb}, Cassandra\citep{lakshman2010cassandra} and DynamoDB\citep{decandia2007dynamo}.

Column-oriented databases store columns of data separately, unlike RDBMSs where data are stored in the form of complete records. They are suitable for vertically partitioned, contiguously stored, and compressed storage systems. Reading of data and retrieval of attributes in such systems is quite fast and less resource intensive than RDBMSs, as only the relevant column is accessed and concurrent process execution is performed for each column \citep{adabi2018optimizing}. Column-oriented databases are highly scalable and Eventually Consistent. Examples of Column-oriented DDMSs are HBase \citep{george2011hbase} and HyperTable \citep{khetrapal2006hbase}.

Document-oriented database are similar to key-value DBs and store data in the form of key and value as reference to a document (i.e., a file). However, document databases support more complex queries and hierarchical relationships. This data model typically uses the JSON format and offers very flexible schema \citep{chodorow2013mongodb}. Although the storage architecture is schema-less for structured data, indexes are well defined in document-oriented databases. SimpleDB is the only database that does not offer explicitly defined indexes \citep{cattell2011scalable, calil2012simplesql}. Document-oriented databases extract metadata to be used for further optimization and store it as documents. CouchDB \citep{anderson2010couchdb} and SimpleDB \citep{chaganti2010amazon} are two examples of Document-oriented DBs.

Graph databases are extensions of Key-value databases. As such, each observation (row) is stored as a dictionary or a series of nested dictionaries (primarily in JSON format). The nested dictionaries contain relational structure. Graph databases offer persistent storage of objects and relationships and support simple and understandable queries with their own syntax \cite{iordanov2010hypergraphdb}. This allows data to be linked together directly, which can be accomplished with one operation making querying more efficient. Modern enterprises are expected to implement graph databases for their complex business processes and interconnected data, as this relational data structure offers easy data traversal \citep{developers2012neo4j}. The most common Graph DB is Neo4J \citep{developers2012neo4j}.

Finally, NewSQL is an emerging DDMS technology that extends NoSQL approaches while building upon attractive features of traditional RDBMSs. Whereas NoSQL does not provide ACID guarantees for database transactions, NewSQL approaches do. As a result, NewSQL approaches combine the best of traditional RDBMSs and NoSQL approaches. At the same time, NewSQL are rapidly evolving and do not always have extensive support. As a result, we mention them as an avenue of considerable potential, but which remain in development and an interest for research\cite{chen2016promises, stonebraker2012newsql, grolinger2013data}. The most popular NewSQL frameworks are NuoDB \citep{brynko2012nuodb}, VoltDB \citep{stonebraker2013voltdb}, Google Spanner \citep{corbett2013spanner} and CockroachDB \citep{corbett2013spanner}.

\subsection{Opportunities for Unstructured Transport Data}
Evolving transport systems ingest data in the formats of images, videos, audio and various other unstructured data formats. As a result, ITS architectures need schema-free databases to store non-related data provided by traffic surveillance and traffic sensor systems, which hitherto could not be stored in traditional RDBMSs. Orru et al.~\cite{orru2017demonstration}, however, built an ITS application with a backend of a NoSQL database to create a public access of public transport information (of GTFS files) all over the world and also search for geotagged photos. NoSQL systems allow for the storage of such schema-less files, which would be difficult to implement in a traditional database. Typically, travel mobility datasets are designed with varying questions (i.e. fields) based on the purpose of the survey that can contain unstructured formats like audio and images. NoSQL databases allow for the efficient storage of travel mobility data. Vela et al.~\cite{vela2018using} focused on the design and storage of accessible transport routes, obtained by means of crowd-sourcing techniques, in a NoSQL graph-oriented database. The authors adopt a graph NoSQL database to address the integration of accessibility data from three sources, namely; existing open data, private data concerning actual accessible routes obtained through crowd-sourcing, and data from existing traffic sensors. NoSQL databases embrace the capability of a seamless integration of varying and non-related data, which is common in transport systems.

\section{Challenges and Opportunities in Processing} \label{sec:processing}
The third challenge identified in the literature is being able process all of the BTD. This concerns the v-word, ``velocity." While processing is required in the management of data (i.e., storage), the main processing challenge is making use of collected data. The methods used to process data are a function of how quickly the processing is required, i.e. whether information is required in real-time or not. There are in general, two approaches to processing BTD: Batch (\textit{ex post}) Processing, and Stream (real-time) Processing. These approaches require implementation using different Processing Engines, or Frameworks. Below we describe the approaches as well as the most common implementing Frameworks.

\subsection{Batch Processing} \label{sec:batchProcessing}
Batch processing is the processing of large, complete, static or historical data sets, and provides information after the entire dataset has been collected \citep{moniruzzaman2013nosql, ji2012big, chen2014data}. In other words, results are not provided in real-time. As an example is OD surveys are conducted until completion of data collection before processing of data aggregation is done.

This approach is mostly adopted when processing finite (or bounded) datasets that are complete, whose size can be estimated, and that are persistently stored on a hard drive. That is, the dataset is unchanging when it is analyzed and includes information for a given period of time (e.g. data from a regional OD survey). The data needs to be complete because the types of calculations done on them require having all of the relevant data, such as when calculating totals and averages. In such situations, datasets must be treated holistically instead of as a collection of individual records. Also, the operations require that the dataset be unchanged for the duration of the calculations. Most common framework for batch processing is Apache Hive \citep{thusoo2009hive}.

\subsection{Stream Processing}
Whereas Batch Processing requires datasets to be complete and static, Stream Processing systems operate on data immediately as it arrives \citep{ranjan2014streaming, chen2014data}. As such, the data being processed does not need to be complete or static. Moreover, the size of the ``entire" dataset is unknown at any given time until data is no longer collected, i.e. it is ``infinite," and its size is irrelevant for Stream Processing. To understand Stream Processing, it is useful to understand Stream Processing workflow.

Typically in a Stream Processing environment, data is received continuously (although not necessarily at a continuous rate), and the data contains information that is not required for the immediate analysis for which results are sought. As such, a first stage of processing is to retain only data relevant to the processing goal. None of the other data is kept or stored. Once data is filtered, processing operations are done on individual observations ``one at a time."  Stream Processing is well-suited to situations in which results are required in real-time. 

An excellent example of situations requiring Stream Processing is Uber, the peer-to-peer ridesharing company \citep{uber2018online}. Uber needs to analyze the location of its riders and to match them with the nearest drivers. They also need to determine the most efficient itinerary for the driver to the rider's origin and destination once picked-up. Moreover, information on the location of the driver needs to be provided to the waiting rider. Once a trip is completed, Uber needs to calculate the cost of the trip and send this information to the rider. All of this requires processing to be done in real-time. An emerging technology for which Stream Processing is already required, and for which it will be required in greater amounts in the future is that of Autonomous Vehicles. While Uber needs to be able to process streamed data quickly, Autonomous Vehicles need to process information (read in data, react) instantaneously.

As with Batch Processing, specialized Processing Frameworks are required for Stream Processing. Also, as with Batch Processing, many such frameworks exist, with the most common being Apache Storm \citep{storm2013storm}, Kafka \citep{thein2014apache} and S4 \citep{neumeyer2010s4}.

\section{Challenges and Opportunities in Cyber-Security}
The fourth challenge relates to the fact that BTD infrastructure needs to be secured from unauthorized access by an attacker. This challenge is related to ensuring transportation system components are securely protected to avoid vulnerabilities exposed for an adversary to exploit and also protect data as it is transmitted on communication channels. We continue to discuss the context of cyber-security in transportation and known vulnerabilities to be considered.

\subsection{Cyber-Security of BTD}
Recent dominance of high-resolution information gathering devices (i.e. Cameras, transponders, wireless routers) and social systems are on a path of fully connectivity known as ``Internet Of Things". A large body of research and standards have evolved on mining rich data ingested by these interconnected devices. Intelligent transport systems, gain access to a wealth of information from interconnected data from GPS location tracking to traffic logs, that aid in public safety, disaster recovery and emergency response. As modern transport devices contain a network of networks made up of embedded communication methods and scope, issues of cyber-security are raised. 

Whilst discussion on IT Security is a fundamental challenge to core IT implementation and not limited to Big Data implementation, a scope of Cyber-Security is worth considering as it can impact on the veracity (truthfulness) of the data harnessed on large-scale integrations. Cyber-Security protects against illegal or unauthorized access to information sources and their communication channels which can disrupt service availability for interconnected devices. There is a need for devices and generated data to be adequately secured against attacks, vulnerabilities and exploits. Potential vulnerabilities that could be exploited in transportation include unsecure vehicle-to-vehicle communication, unauthorized vehicle data interception, seizure of control systems like brakes or accelerators. As an example, a group of civic hackers deciphered and exposed the bus location system of Baltimore in 2015 \citep{baltimore2018online}. In 2016, San Francisco transit was hacked to give unpaid access to commuters for two days \citep{franciscohack2018online}. It is evident that uncontrolled attacks and vulnerabilities can defame the purpose of intelligent transport systems and incur unforeseen losses that can destroy system implementation. Key vulnerabilities that are of concern in Big Transportation Data implementation are discussed below.

Vulnerabilities of Software Applications: Most common threat to security for Big Transportation Data is exploits undertaken in software libraries and bugs. Software packages and Operating system (or firmware) kernels usually expose vulnerabilities or system bugs that hackers can exploit to gain unauthorized access to control the system. Software updates, patches or fixes are periodically developed by software manufacturers to update known vulnerabilities mostly through automatic system updates. As transportation information systems encompass a wide suite of software components (i.e. web server, database, application framework), it is required system updates from trusted manufacturers are allowed and enforced to ensure a robust secured platform for information share.  

Vulnerabilities of Field Devices: BTD ingest high-volume data from dispersed sensor and pervasive devices which are mostly located in remote areas and far from routine supervision. These remote field devices such as traffic lights, cameras, road counter equipments are often in isolated public places and remain susceptible to tampering. Isolated field devices are vulnerable to tampering thus an adversary, who can alter the physical configuration of devices can compromise a system by gaining illicit access to its information source. It is important a level of surveillance is provided for field devices which are deployed in isolated environments.

Vulnerabilities of Communication Networks: Communication devices create an enabling environment for data exchanges between interconnected devices. Network vulnerabilities are well known within wired and wireless network service. Such vulnerabilities allow an attacker to eavesdrop on data packets which are exchanged in the communication channel. Cellular networks, mostly wireless services, are known to be vulnerable to signal intercepts and other threats. Wi-Fi network vulnerabilities are very common in hacker communities, who gain access and exploit the network including devices that are connected. A network map is a sensitive information to an adversary who might be interested in exploiting a transportation system thus its detail should be treated with high confidentiality. Data Encryption and cryptographic algorithms such as Data Encryption Standard (DES) algorithm, Rivest-Shamir-Adleman(RSA) are applied to data packets to perturb the data content as they are transmitted over network channels. The underlying transport layer is made secured by adopting secured communication protocols such as Transport Layer Security(TLS) and Secure Sockets Layer(SSL) which provides privacy and data integrity between communication nodes.

\section{Challenges and Opportunities in Privacy Protection} \label{sec:privacyProtection}
Until now we have focused on challenges related to defining characteristics of Big Transportation Data, namely the three Vs. The fifth challenge relates to the fact that BTD often contains personal data explicitly, or personal information that could be revealed by combining or analyzing data that is not strictly-speaking personal, i.e., ``Personally Identifiable Information" (PII) \cite{tene2011privacy, schwartz2011pii}. In other words, the challenge is related to ensuring the protection of individual privacy with the use of BTD. This is not a challenge uniquely for BTD, and the challenge of privacy protection in the face of PII has been an issue for a long time (see e.g. Sweeney \citep{sweeney2002k} who experiments identifying personal information by linking voter registration data sets to medical records). As a result, we do not concentrate on the general question of privacy protection with PII as it has received a great deal of former attention (see e.g. \citep{schwartz2011pii, mccallister2010guide}). What is unique about BTD is the large amount of temporally and geographically precise location data that can be collected on people. As such, this discussion focuses on the protection of privacy in the context of what we refer to as ``Personally Identifiable Location Information" (PILI).

An example is given by Anthony Tockar \cite{neustar2018riding}, a summer intern at Neustar, an information-analytics who showed how to extract the exact location and time that celebrities used cabs in New York City extracted from open New York City Taxi and Limousine Commission (TLC) data. By joining the two data sets, Tockar found the cash tips paid by celebrities  \cite{neustar2018riding}.

Transportation planning agencies have had access to both PII and PILI in the past through routine data sources collected for planning purposes such as Origin-Destination surveys. As a result they have used techniques to protect privacy both internally, as well as when such data is shared with third parties such as consultants and academic partners.

With greater amounts of or more detailed information about people, these methods will need to be adapted. Such adaptation is becoming increasingly important with open data policies (see e.g., \cite{montreal2018montreal}), which are becoming more common and which by their nature impose much less control on who and the number of people who have access to potentially identifiable information. An understanding of the techniques used for privacy protection in the context of PII, and available for use with PILI, requires an understanding of underlying data Anonymization Operations. We begin with these and then continue with a description of Anonymization Techniques as they have been applied with PII and how they are applicable to PILI. 

\subsection{Data Privacy and the Need for Anonymization}
Information collected for transportation planning and operations purposes can contain ``microdata," i.e., detailed information on individuals and households (addresses, age, sex, etc.) \citep{ghinita2007fast, cormode2009anonymized}. Data attributes (or variables) that identify individuals are referred to as ``Explicit Identifiers." Attributes that do not explicitly identify individuals or households can, in combination with other attributes, potentially identify record owners uniquely \citep{sweeney2002k}. Such attributes (e.g., zip code, sex, date of birth, etc.) are referred to as ``Quasi Identifiers." While being able to identify individuals is an issue in itself, it becomes even more critical when ``Sensitive Attributes'' (e.g, disease, income, etc.) \citep{cormode2009anonymized} are available. 

Another issue affecting privacy protection and concerns is to whom data is available. To best understand the issues surrounding this, we define what we refer to as the Data Chain of Custody (DCC) that describes how data passes from the individual on whom it is collected to the end user of the data. The Data Chain of Custody is an adaptation of Xu et al.'s \citep{xu2014information} data ``User Roles."

The chain begins with the Data Owner (same term is as Xu et al.) who is the person on whom data is being collected. The Owner's information is recorded by the ``Data Recorder" typically a device, such as a smartphone. The Data Collector arranges the collection, stores and curates the data for the Data Analyst. It can be an individual researcher, a governmental institution (e.g. regional planning authority) or a private company. Data Collectors can collect data for their own purposes, or on behalf of others. Data Analysts process, analyze and integrate collected data for the End User. Multiple roles can be played by the same individual or institution, so that for example the Data Collector might also be the Data Analyst and End User. Sometimes the Data Owner (in the case of Location Based Services) can also be the End User. We quickly provide three examples of BTD and the DCC.

The first example relates to the smartphone travel survey platform, Itinerum \citep{patterson2017itinerumtrip}. This platform allows researchers to develop and administer their own customized smartphone travel surveys (see e.g. \cite{patterson2018designingJUM}). While the platform also allows some data processing, in this example, we assume that the survey administrator only uses it to collect data and does analysis in-house. As such, this example involves a municipality that undertakes a smartphone travel survey that it will use for analysis of their local transportation system, as the City of Montreal did in 2016 \cite{patterson2017mtlisctsc}. In such a circumstance, the Data Owner is the respondent with their smartphone being the Data Recorder. The Data Collector is the Itinerum project that is collecting the data on behalf of the municipality. The municipality peforms analysis on the data and therefore is the Data Analyst. Because the municipality will use the analytical results from the collected data, it is also the End User.

\begin{figure}[h]
	\centering
	\includegraphics[width=1.0\textwidth]{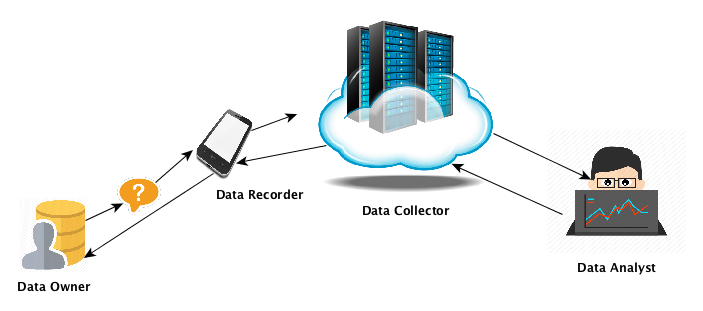}
	\caption{Dataflow across data agents}
	\label{fig:dataflow}
\end{figure}

The second example is a someone requesting a list of nearby restaurants through Google Maps on their smartphone, also known as a Location Based Query. In this case, the Data Owner is the person searching for restaurants and their phone the Data Recorder. Google is the Data Collector since it developed the app and infrastructure and stores the owner's location data. Google is also the Data Analyst since it processes the request and returns a list of nearby to the User. As such, the Owner is also the ``End User". See Figure \ref{fig:dataflow}.

Lopez and Farooq \citep{lopez2018blockchain} propose a transportation blockchain system to protect the personal travel information and improve the privacy of respondents to passively solicited data. The proposed system protect users by making them the data owners and controllers of their personal information and is secured by a private key which can be accessed through smart contracts. The blockchain performs the role as a data collector by assigning keys, maintaining a transactional ledger and smart contracts to the information which the data owner seeks to share. Data Analyst mostly third parties require a smart contract to access travel information.

Data privacy risks are related to the DCC and in particular who, and under what circumstances, has access to the data. A data privacy breach results when someone's identity (possibly associated with Sensitive Attributes) is revealed in a dataset when it is not supposed to be. This can happen unintentionally and with no malicious intent. When it happens intentionally and with malicious intent, it is referred to as ``Adversarial." \citep{hasan2013decentralized, lindell2005secure}

As the number of people accessing data, and the number of people accessing data whose identities are not known, increases, so does the risk of adversarial data privacy breaches. When data is available to few known individuals (e.g. to data analysts in a municipal planning agency), privacy risks are limited. This is because the people with access are known and typically employees operating under regulations. Also, fewer people accessing data implies lower probabilities of discovery of private information that could be revealed when combining data sources, Quasi-identifiers, etc. This situation is at one end of the privacy risk spectrum with open data being at the other.

With Open Data there are unknown numbers of unknown people accessing data. So, the characteristics of data and the degree to which data is available to known or unknown users determines the risk of the revelation of private information. Privacy protection with PII is implemented with a number of different anonymization operations, which are applied in different combinations in Anonymization Techniques (or Anonymization Models). We first discuss Anonymization Operations and then Anonymization Techniques.

\subsection{Anonymization Operations}
The most popular anonymization operations used in application are: generalization, suppression and perturbation. Generalization performs anonymization on data by replacing some values of an attribute with a taxonomy of its parent value \citep{nergiz2008towards, bayardo2005data}. A set of attributed values are replaced by a general categorical description value (e.g. replacing language spoken at home with English or Other). Generalization operations are mostly applied to quasi-identifiers and sensitive attributes, and reduce the probability of uniquely identifying a record owner. A numerical interval or range is typically used to generalize numerical attributes. Specialization is achieved when generalization is reversed by returning the detail of specific values.   

Whereas Generalization works with taxonomies, Suppression also replaces values of an attribute with a special key \citep{aggarwal2005anonymizing, bayardo2005data}, typically an asterisk (*). As data is suppressed, identifiable values are replaced by special keys to make values non-identifiable. Suppression is generally applied to explicit-identifiers and quasi-identifiers. Suppression ensures personal information is not disclosed. 

On the other hand, Perturbation performs anonymization by distorting the original data with the addition of noise, data swapping, value aggregation and generation of synthetic data. Statistical approaches are used to perturb data values \citep{aggarwal2005anonymizing}. Perturbation generally replaces real data values as well so that data does not correspond at all to the original value associated with the individual. When statistical methods are used to perturb data, while attribute values are not those of the original individual, the aggregate characteristics of the attributes are the same as for the entire dataset.

\subsection{General Anonymization Techniques}
Anonymization Techniques use combinations of the Anonymization Operations described above to anonymize PII. The most popular techniques used to limit disclosure of identifiable information are are K-anonymity-based techniques and Differential Privacy. The anonymization techniques address privacy protection under different circumstances of access to data.

\subsubsection{K-anonymity-based Techniques}
K-anonymity-based techniques are relevant in the following data access circumstances. The original dataset is contained in one or multiple tables and all Explicit Identifiers have been removed. K-anonymity requires that after removal of Explicit Identifiers, each record must be indistinguishable from at least another k-1 records with respect to any given quasi-identifier \citep{sweeney2002k, aggarwal2005anonymizing}. For example, when k-anonymized, if a given record has a given value for an attribute, there will be at least k-1 other records with that same value. As such, k-anonymization removes the uniqueness of distinct values for a quasi-identifier through generalization and suppression operations. 

While k-anonymity protects against identity disclosure, it is insufficient to prevent attribute disclosure (being able to associate a unique attribute value to a given record). L-diversity on the other hand is a concerned not so much with identity disclosure, but with the ability to associate Sensitive Attributes to a given record Machanavajjhala et al. \cite{machanavajjhala2006diversity}. An equivalence class (i.e., a set of records that are indistinguishable from each other with respect to a given quasi-identifying attribute) is said to have l-diversity if there are at least l ``well-represented" values for the sensitive attribute. As such, this is fundamentally k-anonymity but for the special case of Sensitive Attributes \cite{machanavajjhala2006diversity}. A table is said to have l-diversity if every equivalence class of the table has l-diversity. As a result, and like k-anonymization, l-diversity removes the uniqueness of distinct values but for a sensitive attribute through generalization and suppression operations.

A common problem of both k-anonymity and l-diversity is that they cannot guarantee the protection of private data if information about the global distribution of an attribute is known, e.g., if someone had access to the entire table containing any given k-anonymized or l-diversified attribute. This problem is particularly acute if the distribution of the attribute in question has few values and/or is highly skewed towards a few values, e.g. if 90\% of the values of tips given to drivers (see example in Section \ref{sec:privacyProtection}) in a given dataset were 0, it would be straightforward to infer that a given individual did not leave tips. To address this problem, the t-closeness anonymization technique has been developed.

T-closeness \citep{li2007t} itself is a measure of the degree to which a distribution is skewed towards a few values. As t-closeness increases, a distribution becomes more skewed towards a few variables. The t-closeness technique amounts to adjusting the distribution of sensitive attributes to assure that the global distribution does not have few values and is not highly skewed towards any, or a few, of those values. An equivalence class is said to have t-closeness if the distance between the distribution of a sensitive attribute in this class and the distribution of the attribute in the whole table is no more than a given threshold \textit{t}. A table is said to have t-closeness if all equivalence classes have t-closeness. T-closeness is ensured through generalization and suppression operations.

\subsection{Differential Privacy}
Strictly speaking, Differential Privacy is not a technique, but rather a property of the anonymization process. The concept of Differential Privacy was originally introduced by Dwork et al. \citep{dwork2008differential} and is relevant in the following data access circumstances. There is an original database ($D$) with Explicit- and/or Quasi-identifiers and Sensitive Attributes. There are also two agents accessing the data either indirectly or directly. The data User wants to learn about the characteristics of the original dataset by making queries to it, but does not have direct access to the original data. The Curator (a software component between other software layers, or middlewear) has direct access to the original data, but has the role of modifying it thus creating a new dataset ($D^\prime$) to which the User has direct access.

Critical to understanding Differential Privacy is the notion of Privacy Degradation. Privacy Degradation describes the fact that as queries are made to a database, the results of each additional query provide information that can be compared with previous results. As such, it is possible, all else equal, to learn about individual observations in a modified database by comparing results made with different queries.

Ultimately, the Curator's role is two fold. First, remove Explicit identifiers from the original database and perform modifications (Perturbations) on the Quasi-identifiers and Sensitive Attributes. These perturbations are typically created by adding noise drawn from a Laplace distribution to Quasi-identifiers and Sensitive Attributes. It is important to note that $D\prime$ itself is dynamic, so that it might not be the same for subsequent queries from the User. The degree to which $D\prime$ is different from $D$ is referred to as epsilon ($\epsilon$). With Differential Privacy $\epsilon$ is also dynamic and is a function of the number of queries from the user.

\subsection{Location Privacy}
Anonymization methods discussed so far have been developed and applied primarily to PII \citep{mccallister2010guide, schwartz2011pii} data. The large amounts of temporally and spatially precise BTD can be thought of as Quasi-identifying data, but the techniques mentioned above are not suitable to ensure privacy protection with this PILI data. There are two broad categories of circumstances under which anonymization of PILI can take place. The first relates to when data is transferred from the Data Recorder to the Data Collector. The second is when data is transferred from the Data Collector to the Data Analyst. The first category is referred to as Location Based Query (LBQ) anonymization \citep{kalnis2007preventing, ghinita2008private}. This might happen for example if the true location of the Data Owner is anonymized or obscured by the Data Recorder before being sent as part of an LBQ, such as a search for nearby restaurants. The second category is when data is transferred from the Data Collector to the Data Analyst. While the first type of anonymization is important, we believe it to be less of a challenge to the use of BTD than the second. This is because with LBQs the Data Collector and Analyst will typically be known and presumably trusted if the Data Owner is willing to share their information with them. Of greater practical concern is what happens as data is transferred from the Collector to the Analyst, since the identity of analysts may not be known, and there may be many, particularly in the case of Open Data. As a result, in this paper we concentrate on techniques relevant for anonymization that takes place between the Data Collector and the Data Analyst, i.e. to data ``publishing."

There are four main techniques available for location anonymization appropriate for PILI data when it is published. The techniques differ along three dimensions: whether a Data Owner ID persists or not; the approach used for obfuscating location; and whether or not the anonymization is done real time.

\subsubsection{Spatial Cloaking}
Spatial Cloaking \citep{gruteser2003anonymous, chow2006peer} is used when data is static (i.e. when it has already been recorded and stored). When location data are spatially cloaked the Data Owner IDs persist across observations, but locations reported to the Data User are adjusted. In particular, and instead of providing the original location data (i.e., latitude and longitude), the data are spatially aggregated so that the Data User is provided a spatial buffer known as the Anonymized Spatial Region (ASR) \citep{terrovitis2008privacy}. The size of the buffer is dynamic and is a function of the number of other Data Owners on whom data is reported. In particular, the ASR is large enough to encompass the data of at least k-other Data Owners. As such, it can be seen as a spatial k-anonymization. Since ASRs are dynamic this technique is also computationally intensive. This technique could be used with trip-end location data or trajectory data.

\subsubsection{Mixed Zones}
Mixed Zone (MZ) \citep{beresford2004mix} anonymization is used when data is static. With MZ-anonymization, it is Data Owner (pseudonym) IDs that are obfuscated and not their locations. This is done first by defining zones through which the Data Owner passes. As the Data Owner passes through zones, their IDs are modified so that it is not possible follow an individual as they pass through the different zones. Mixed Zones is a more general approach that encompasses the special case of the Vehicular Mix Zone approach.

\subsubsection{Dummy Trajectories}
As with Spatial Cloaking and Mixed Zones, the Dummy Trajectories technique \cite{you2007protecting} is used to anonymize static spatial data, particularly static trajectory data. As with Spatial Cloaking, Data Owner (pseudonym) IDs persist throughout the data and unlike MZ does not involve the creation of zones. This method amounts to perturbing location data over the course of a trajectory. 

\subsubsection{Path Confusion}
As with Spatial Cloaking and Dummy Trajectories with Path Confusion \citep{hoh2005protecting} Data Owner IDs persist. Like the Dummy Trajectory approach, the location data is perturbed directly. Unlike these other methods, the data in question is not static but is arriving in real time. The key concern with this approach is to make it impossible to predict a future location based on the dynamic data. As such, this is a more statistically involved approach that not only perturbs location data directly, but also associated bearing and speed data. Due to the the statistical complexity and the need to treat each data point in real time, it is computationally intensive.

\section{Cross-Cutting Opportunities and Challenges} \label{crossCutting}
The previous sections have focused on the primary challenges facing the widespread use of BTD and the opportunities to overcome these challenges. The opportunities in these sections have included those that are applicable to one of the challenges at a time. In this section we discuss opportunities (and challenges associated with their implementation) that will help in overcoming more than one of the ``3-V" challenges. In one way or another the approaches required to overcome the 3-V challenges amount to being able to add computing resources, constrained ultimately by hardware. Cloud computing \citep{armbrust2010view} involves adding resources virtually. That is to say that instead of adding physical resources (e.g. servers), it is possible to add resources through software that mimic the behaviour of physical hardware. This can be done ``privately" on infrastructure managed directly or ``publicly" by going through Cloud computing providers such as Amazon Web Services (AWS), Google Cloud, Microsoft Azure, Rackspace, etc. Cloud computing allows the possibility to quickly add resources, and thereby scale systems in near real time and even automatically. Using Cloud resources reduces the requirements for internal expertise and allows granular addition of resources where the addition of physical infrastructure is ``lumpier."

The costs of using Cloud computing services need to be traded-off with the costs of managing physical infrastructure, but is becoming increasingly competitive for almost all typical computing requirements. It is likely to become even more competitive over time making the choice of using Cloud computing somewhat easier on a cost-only basis. Another issue with Cloud computing, however, is the loss of control over where data is physically stored (i.e., where physical servers are located). This can be an issue for transportation authorities that have traditionally operated under circumstances where all data is stored ``internally." Of course, Cloud computing can be done ``privately" although this still requires a great deal of internal resources (more than managing infrastructure directly) and is likely only viable for large organizations. 

Cloud resources \citep{bhardwaj2010cloud} can be added through three service models: Infrastructure as a service (IAAS), Software as a service (SAAS) and Platform as a service (PAAS). IAAS is the most direct model for adding additional resources. It involves the addition of virtual infrastructure (e.g. computers) that are managed by the service user. As such, software required by the user is installed and maintained on the additional virtual resources. SAAS is the most limited model with users subscribing to particular application software and databases. Microsoft Office online, SQL Server web, ArcGIS online are all examples of this. PAAS is the most involved of the three models. PAAS solutions are designed primarily for technology developers and as a result provide all necessary elements of a development environment. That is PAAS comes with pre-packaged operating system, web server, database and programming languages. PAAS examples include IBM Cloud, Microsoft Azure, Blockchain \citep{lopez2018blockchain}, etc.

\section{The Future of BTD in Transportation}
The rapid emergence of different tools for data collection has led to an unprecedented potential not only to collect, but to integrate data from many sources and potentially to revolutionize how transportation planning and operations are done. This potential has not been lost to transportation researchers, but current research has focused on techniques for collecting data or on inferring relevant transportation information from this data. While critical to fulfilling the potential, we define four existing, higher-level challenges and opportunities to the large-scale use and integration of BTD for planning and decision making purposes. Three of the challenges (and opportunities) are related directly to the 3-Vs of Big Data more generally. A fourth relates to all of the 3-V challenges collectively, and a fifth concentrates on the challenge of privacy protection. This is particularly relevant to BTD due to the large amount of temporally and spatially precise data collected. In our view, BTD will not be able to fulfill its promise if these challenges are not met. We consider the challenges related to the 3-Vs (Volume, Variety and Velocity) first and continue with those related to privacy.

\subsection{Challenges Associated with the 3-Vs}
The challenge associated with the sheer Volume of BTD that are, and that will continue to be, available in ever-larger quantities will continue to place pressure on traditional vertically-based Database Management Systems. The ability to vertically scale these systems is already at its limits and as a result the future will increasingly (and perhaps eventually entirely) require horizontally scaled systems deployed using distributed architectural approaches. While current approaches are dominated by CA architecture, there is a gradual drift towards AP architecture ensuring high Availability, Fault Tolerance and Delayed or Eventual Consistency. This pattern will continue into the future and AP architectures are likely to become the dominant approach in the near, and for the foreseeable, future.

While large Volumes of data present their own challenge, being able to process data coming in at different rates and increasingly in real time is the challenge of Velocity. Traditional Batch Processing methods are ill-adapted to the onslaught of real-time data that also needs to be processed in real time. As a result, in the future the need to increasingly devote resources to Stream Processing methods will become more prominent. While Stream Processing will undoubtedly make up a larger proportion of processing, Batch Processing, when appropriate will continue to play an important role. Batch Processing will remain the mainstay of processing for static datasets and analysis requiring access to a finite dataset. In the future, processing will not simply take place as Batch or Stream processing, but is likely to involve techniques that take advantage of both approaches, such as emerging ``Lambda" architectures \citep{lambda2016online}.

The Variety (different data and file formats from different sources) of BTD represents another key challenge. Traditional structured Relational Database Management Systems that require defined data schemas are incapable of handling and integrating data from different sources; something necessary but also which provides one of the most important aspects of the potential of BTD. As such, the move away from traditional RDBMSs and towards more flexible non-relational DB systems will need to continue to cope with the many different formats. The most common production-ready flexible systems are NoSQL-based and such systems are set to become more commonplace and the \textit{de facto} standard in the near future. At the same time, new approaches are already evolving to overcome the constraints of NoSQL systems and in particular new flexible systems that are also ACID compliant with NewSQL-systems being the most likely to replace NoSQL.  

Finally, Cloud computing will be key to meeting all of the 3V challenges. It will provide the possibility to granularly, quickly and automatically add computing resources necessary to cope with increased Volumes, Velocity and Variety of data. The economic case for Cloud computing seems undeniable, but its use will likely involve the necessity to give up the ability to store data internally for most organizations. As such, in order for it to play its facilitating role in allowing BTD to revolutionize planning, organizations will need to be convinced that collected data is stored sufficiently safely. This will likely happen through a combination of attempts on the part of Cloud service providers to convince organizations of the safety of data and an eventual institutional acceptance of using these services.

\subsection{Challenges Associated with Cyber-Security and Privacy}
The last major challenge is that of security and privacy. The first three challenges are essentially technical in nature and if not met, it will simply not be possible to take advantage of the potential of BTD. Privacy on the other had is both a technical, as well as social/political challenge. The social/political challenge is that of Data Owners (the public) being willing to share their data with Data Collectors and subsequently to Data Users. Ensuring this willingness has three elements. The first is that related to security, which is a challenge facing all IT. Network threats have not been dominant in the transport industry as compared to other sectors. Notwithstanding, there is a rising need to build robust and secured transportation infrastructure that is protected from wide range of system vulnerabilities and exploits. As a step to improve security, network threats to existing systems need to be assessed and reported. Network assessment tools (e.g. Wireshark\citep{orebaugh2006wireshark}) have become popular for network monitoring to gain better visibility of vulnerability to cyber threats. Enterprise architectures deploy network security systems such as firewalls and proxy servers, which monitors incoming and outgoing network traffic by adhering to strict security rules.  A final step to achieve secured computing is to improve communication between trusted cyber-security experts and operators at national and local transportation agencies. Such communication notifies on active security threats and provide relevant information on managing such threats. 

The second relates to the knowledge of Data Users with respect to the nature of their data that is being shared as well as with whom. This has been prominent lately with the necessity for companies to comply with European GPDR regulations. We believe an important challenge related to this is also in the simple and clear explanation to the Data Owners of what data is being collected and shared, something not easily accessible through typical Terms of Use and Consent Forms. The third, is that related to privacy protection and more specifically privacy protection in the context of ``published" data. That is, data that is shared with Data Users. Traditionally data was published to relatively few people whose identities were known. The advent of Open Data has resulted (and will increasingly result in the future) in many more people whose identities are not known having access to published data. Moreover, with the anonymity of Data Users, the risk of the adversarial use of such data, particularly with increasing ``background'' knowledge, and therefore the threat of privacy breaches will only increase.  As such, ensuring willingness on the part of Data Owners will increasingly involve assurances around the protection of privacy from collected data. These assurances will be based upon methods of anonymization. As a result, anonymization is critical to ensure the trust of Data Users.

There are already many anonymization techniques that have been developed for the purposes of privacy protection with both tabular as well as geographic data, and this is a lively area of academic and private sector research. At the same time, this is an evolving field and one that will have to continue to evolve as more data becomes open. The primary reason for this is the growth of Open Data for two reasons. First, as more data become open, there will be more people able to access it anonymously and as a result a greater threat of adversarial use of the data. Compounding is the fact that as more data becomes open, more ``background" knowledge will also become open further expanding the threat of potential privacy breaches. As a result not only will it be necessary for anonymization techniques to evolve, but caution related to the data that is made open will need to be taken.

\section{References}
\bibliographystyle{plainnat}
\bibliography{sample.bib}

%
%
%
%

\end{document}